\documentclass[12pt,reqno]{article}
\usepackage{amssymb,amsmath,amsthm}
\usepackage{a4}

\newcommand{\tup}[2]{(#1_1,\ldots,#1_{#2})}
\newcommand{\tupp}[3][1]{#2_{#1},\allowbreak\ldots,\allowbreak #2_{#3}}
\newcommand{\clo}[2][{}]{\mathrm{cl}_{#1}(#2)}
\newcommand{\inte}[2][{}]{\mathrm{int}_{#1}(#2)}
\newcommand{\Rset}{\mathbb{R}}   

\newtheorem{lem}{Lemma}
\newtheorem{thm}{Theorem}
\newtheorem*{thm*}{Theorem}
\newtheorem{claim}{Claim}

\title{Expressing the cone radius in the relational calculus with real
polynomial constraints}
\author{Floris Geerts\\
University of Limburg (LUC)\\ Dept. WNI\\ B-3590 Diepenbeek, Belgium\\
\texttt{floris.geerts@luc.ac.be}}
\date{}

\begin{document}

\maketitle
\begin{abstract}
We show that there is a query expressible in first-order logic
over the reals that returns, on any given semi-algebraic set $A$,
for every point a radius around which $A$ is conical.  We obtain
this result by combining famous results from calculus and real
algebraic geometry, notably Sard's theorem and Thom's first
isotopy lemma, with recent algorithmic results by Rannou.\\

\end{abstract}
\section{Introduction}\label{sec:intro}

The framework of constraint databases, introduced by Kanellakis,
Kuper and Revesz~\cite{kkr_jcss95}, provides a nice theoretical
model for spatial databases~\cite{pvv_pods94}. A spatial dataset
is modeled using real polynomial inequality constraints; such sets
are also known as semi-algebraic
sets~\cite{Benedetti,bcr_real98}. The relational calculus
(first-order logic) with real polynomial constraints then serves
as a basic query language, denoted here as FO\@.

The study of the expressive power of query languages for
constraint databases is an active domain of
research~\cite{cdbook}. One of the problems in particular that
received attention in recent years is that of determining the
exact power of FO in expressing topological properties of spatial
databases~\cite{bb2000,gk_dbpl99,groseg00,kpv_jsl99,seggie}. One such
property, which is central in this research, is that locally
around each point, a semi-algebraic set has the topology of a
cone. A radius at which this behavior shows is called a \emph{cone
radius} around the point for the set.

Accordingly, a cone radius query is a query that returns, for a
semi-algebraic set $A$, a set of pairs $(\vec{p},r)$ giving for
every point $\vec{p}$ a cone radius $r$ in $\vec{p}$ for $A$.
In this paper, we show that there exists an FO formula expressing
a cone radius query. So far, this was only known in two
dimensions~\cite{gk_dbpl99}. Expressibility of the cone radius,
apart from being a natural question in itself, has also
applications: for example, it has been linked to the
expressibility of piecewise linear approximations~\cite{Geerts2}.

\section{Preliminaries}\label{sec:prelim}
\subsection{Spatial databases and Queries}

A \textit{semi-algebraic set in} $\Rset^n$ is a finite union of
sets definable by conditions of the form
$$
f_1(\vec{x})=f_2(\vec{x})=\cdots=f_k(\vec{x})=0,\,
g_1(\vec{x})>0,\, g_2(\vec{x})>0,\, \ldots,\, g_\ell(\vec{x})>0,
$$
with $\vec{x}=\tup{x}{n}\in\Rset^n$, and  where
$f_1(\vec{x}),\ldots,f_k(\vec{x}),g_1(\vec{x}),\ldots,g_\ell(\vec{x})$ are  multi-variate polynomials in the variables
$x_1,\ldots,x_n$ with real coefficients. A \textit{database schema}
$\mathcal{S}$ is a finite set of relation names, each with a given
arity.
 A \textit{database} over $\mathcal{S}$ assigns to each $S\in\mathcal{S}$
 a semi-algebraic set $S^D$
in $\Rset^n$, if $n$ is the arity of $S$. A \textit{$k$-ary query
over $\mathcal{S}$} is a function mapping each database over
$\mathcal{S}$ to a semi-algebraic set in $\Rset^k$.

\noindent
As query
language we use \textit{first-order logic} (FO) over the
vocabulary \mbox{$(+,\cdot,0,1,<)$} expanded with the relation names in
$\mathcal{S}$. A  formula $\varphi(x_1,\ldots,x_k)$ expresses the
$k$-ary query defined by $$ \varphi(D):=\{\tup{a}{k}\in\Rset^k\mid \langle
\Rset,D\rangle\models\varphi(a_1,\ldots,a_k)\}, $$ for any
database $D$. Note that $\varphi(D)$ is always semi-algebraic
because all relations in $D$ are; indeed, by Tarski's theorem~\cite{vandendries_jsl88}, the
relations that are first-order definable on the real ordered field
are precisely the semi-algebraic sets.

An example  of a query expressed in FO is the following: Let
$\mathcal{S}$ be a schema containing the relation name $S$.
Consider the FO-formula
$$
\varphi_{\text{int}}(\vec{x}):=(\exists \varepsilon
>0)(\forall x_1')\cdots(\forall
x_n')(\|\vec{x}-\vec{x}'\|<\varepsilon\rightarrow
S(x_1',\ldots,x_n')). $$

For any database $D$,  $\varphi_{\text{int}}(D)$ equals  the interior
of $S^D$. However,  not  every query is first-order expressible:
the query which asks whether a set is connected is
not expressible in FO\@. This result and other results related to
constraint databases have recently been collected in a single volume
\cite{cdbook}.
\subsection{Cones}\label{subsec:cones}

Let $A\subseteq\Rset^n$ be a semi-algebraic set and
$\vec{p}\in\Rset^n$ a point not in  $A$. We define the
\textit{cone with base $A$ and top $\vec{p}$} as the union of all
closed line segments between $\vec{p}$ and points in $A$. We
denote this set with
$\mathrm{Cone}(A,\vec{p}):=\{t\vec{b}+(1-t)\vec{p}\mid \vec{b}\in
A,\ 0\leqslant t \leqslant 1 \}$. For a  point $\vec{p}\in\Rset^n$
and $\varepsilon>0$, denote the closed ball centered at $\vec{p}$
with radius $\varepsilon$ by  $B^n(\vec{p},\varepsilon)$, and
denote the sphere centered at $\vec{p}$ with radius $\varepsilon$,
by  $S^{n-1}(\vec{p},\varepsilon)$. The following well-known theorem
says that, locally around each point, a semi-algebraic set has the
topology of a cone.

\begin{thm*}[\cite{Benedetti,bcr_real98}]
 Let $A\subseteq\Rset^n$ be a semi-algebraic set
and $\vec{p}$ a point  of $A$. Then there
is a real number $\varepsilon>0$ such that the intersection $A\cap
B^n(\vec{p},\varepsilon)$ is homeomorphic to the set
$\mathrm{Cone}(A\cap S^{n-1}(\vec{p},\varepsilon),\vec{p})$.
\end{thm*}
\noindent Any real number $\varepsilon>0$ as in the lemma is
called a \textit{cone radius} of $A$ in $\vec{p}$.

Let $\mathcal{S}$ be a schema containing a relation name $S$ of arity $n$.
A \textit{cone radius query} $Q_{\text{radius}}$ is a query
which maps any database $D$ over $\mathcal{S}$ to a set of pairs $(\vec{p},r)\in\Rset^n\times\Rset$
such that for every point $\vec{p}\in S^D$ there exists at least one pair 
$(\vec{p},r)\in Q_{\text{radius}}(D)$, and  for every 
$(\vec{p},r)\in Q_{\text{radius}}(D)$, $r$ is a cone radius in $\vec{p}$ for $S^D$.

We will use the following notation:
Let $A\subseteq B\subseteq\Rset^n$, the closure of $A$ in $B$ is denoted by $\clo[B]{A}$, and
$\inte[B]{A}$ indicates the interior of $A$ in $B$. When the
ambient space $B$ is $\Rset^n$,  we omit the subscript
$B$. We denote $\clo{A}-A$ (the frontier of $A$) with
$\partial A$.
\subsection{Whitney stratification}\label{subsec:Whitney}

Every semi-algebraic set $A\subseteq\Rset^n$ can be ``nicely''
decomposed in a finite sequence $\mathcal{Z}$
 of $n+1$ semi-algebraic sets $Z_0,\ldots,Z_n$, called \textit{strata}, with the following
 properties. For each $i=0,\ldots,n$:
\begin{enumerate}
\item $Z_i$ is either  a $C^1$ semi-algebraic set in
$\Rset^n$ of dimension $i$, or an empty set; and
\item each triple $(Z_i,\vec{p},Z_j)$ for $i<j$ and $\vec{p}\in Z_i$ has the
\textit{Whitney property}.
\end{enumerate}
Such a  decomposition is called a $C^1$-\textit{Whitney
stratification of $A$}. We will  not need the precise definitions of when
a set is $C^1$ and of the Whitney property.
We refer to the paper of Rannou~\cite{Rannou} for more details.

We remark that in this paper we do not require the
\textit{frontier condition}, which says that the frontier of a
stratum is the union of lower dimensional strata, and also do not suppose strata to
be connected. Both properties are not necessary for Thom's first
isotopy lemma~\cite{Gibson,Shiota}, which we will use in our proof
in Section~\ref{proof}.

\section{Constructing a Whitney stratification}\label{sec:stratification}

Let $A$ be a semi-algebraic set in $\Rset^n$.  We shall construct
a $C^1$-Whitney stratification $\mathcal{Z}$ of the closure $\clo{A}$ such that
$A$ is the union of connected components of strata of
$\mathcal{Z}$. We then say that $\mathcal{Z}$ is \textit{compatible with
$A$}. This construction will be expressible in FO\@.

Our construction is an adaptation of the construction given by
Shiota~\cite[Lemma {I}.2.2]{Shiota}.

We define $Z_n$ as the subset of $A$ where $A$ is locally $C^1$
and of dimension $n$. Now suppose
that the strata $Z_n,\ldots,Z_{k+1}$ have already been
constructed. Then the stratum $Z_k$ is constructed as follows.
Define $A_0=A$ and $A_1=\partial A$. For $i=0,1$ construct
\begin{gather}
\begin{split}
R^i_k:=\{\vec{p}\in A_i-\bigcup_{j=k+1}^n Z_j \mid \text{$A_i$ is
$C^1$ and of dimension $k$}\\\text{in a neighborhood of $\vec{p}$}\}
\end{split}
\label{een}
\\
W^i_k:=\bigcap_{j=k+1}^n\inte[R^i_k]{\{
\vec{p}\in R^i_k\mid \text{$(R^i_k,\vec{p},Z_j)$ has the Whitney
property}\}}\label{twee}
\\
Z_k^i:=W^i_k-\clo{R^{1-i}_k}.\label{drie}
\end{gather}
Then we define $Z_k:=Z_k^0\cup Z_k^1$.

The stratum $Z_k$ has indeed the desired properties: By (\ref{een}) it is
$C^1$ and of dimension $k$, (\ref{twee}) guarantees that for all points
in
$Z_k$, and for any $j>k$,
the triples $(Z_k,\vec{p},Z_j)$  have the Whitney
property, and (\ref{drie}) ensures that the connected components of $Z_k$
lie either completely in $A$ or completely in $\partial A$.

It is well known~\cite{DriesMiller,Shiota}  that the dimension of
$A_i-\bigcup_{j=k}^n Z_j$ is strictly less
than the dimension of $A_i-\bigcup_{j=k+1}^n
Z_j$ for $i=0,1$. Hence, the stratification $\mathcal{Z}$ will consists of
exactly $n+1$ strata $Z_k$, some of which may be empty.

We now show that the above construction is in FO\@.
\begin{thm} Let $\mathcal{S}$ be a database schema containing  a
 relation name $S$ of arity $n$. The $n$-ary query
$Q_{\text{$k$-stratum}}$, which takes as input a database $D$ over
$\mathcal{S}$, and returns the $k$th stratum $Z_k$ of
the stratification $\mathcal{Z}$ constructed above for $A=S^D$, is expressible in FO\@.\label{thm_wfo}
\end{thm}

In order to prove FO-expressibility of these queries, it
is sufficient to show that the sets $R^i_k$, $W^i_k$, and $Z_k^i$
occurring in the construction are FO-expressible. But this follows 
immediately from the work of Rannou~\cite{Rannou}. Indeed, from that work
we can deduce the following lemma, which immediately implies Theorem~\ref{thm_wfo}:
\begin{lem} (i)
Let $\mathcal{S}$ be a database schema containing  a relation name $S$
 of arity $n$. The $n+1$ queries of arity $n$, defined as
\begin{equation*}
\begin{split}
Q_{\text{k-reg}}(D):=\{\vec{x}\in\Rset^n \mid \vec{x}\in S^D \land
(\text{$S^D$ is $C^1$ and of dimension $k$}\\\text{in
an open neighborhood of $\vec{x}$})\},
\end{split}
\end{equation*}
for any  database $D$ over $\mathcal{S}$, are all expressible
 in FO\@.\\
(ii) Let $\mathcal{S}$ be a  database schema containing two
relation names $S_1$ and $S_2$ of arity $n$.
The $n$-ary query, defined as
\begin{equation*}
\begin{split}
Q_{\text{Whitney}}(D):=\{\vec{x}\in\Rset^n\mid
S_1^D,S_2^D \text{ are $C^1$ and $(S_1^D,\vec{x},S_2^D)$}\\
\text{has the Whitney property}\},
\end{split}
\end{equation*}
for any  database $D$ over $\mathcal{S}$, is expressible in FO\@.
\end{lem}

\section{Expressing the cone radius in FO}\label{proof}

We are now ready to prove the main result of this paper.
\begin{thm}\label{thm:cones} 
There exists an FO-expressible cone radius query.
\end{thm}
\begin{proof}
Consider a semi-algebraic set $A$ in $\Rset^n$, and let
$\mathcal{Z}$ be the $C^1$-Whitney stratification of $\clo{A}$
compatible with $A$. Let $\vec{p}\in\clo{A}$ and define the $C^1$-map
$$
f_{\vec{p}}:\clo{A}\rightarrow\Rset:\vec{x}\mapsto\|\vec{x}-\vec{p}\|^2.
$$
We will need Thom's First Isotopy Lemma~\cite{Shiota}. Applied to the $C^1$-map
$f_{\vec{p}}$ and the $C^1$-Whitney stratification $\mathcal{Z}$,
this lemma can be stated as follows: For any $a<b$,
\begin{enumerate}
\item[(a)] If $f_{\vec{p}}$ is  proper, i.e.,
$f^{-1}_{\vec{p}}([a,b])$ is compact, and 
\item[(b)] if for each stratum $Z\in\mathcal{Z}$, the restriction $$
f_{\vec{p}}|(Z\cap \inte{B^n(\vec{p},b)-B^n(\vec{p},a)})\rightarrow
(a,b)\subseteq\Rset$$
has no critical points (this will be explained later),
\end{enumerate}
 then for any $c\in(a,b)$, there exists a homeomorphism
$$
h:\clo{A}\cap\inte{B^n(\vec{p},b)-B^n(\vec{p},a)}\rightarrow
(\clo{A}\cap S^{n-1}(\vec{p},c))\times (a,b).
$$
Moreover, this homeomorphism satisfies the
following two properties:
\begin{enumerate}
\item[(i)] For each $d\in(a,b)$, $h(\clo{A}\cap S^{n-1}(\vec{p},d))=(\clo{A}\cap
S^{n-1}(\vec{p},c))\times\{d\}$, and
\item[(ii)] $h(Z\cap
\inte{B^n(\vec{p},b)-B^n(\vec{p},a)})=(Z\cap S^{n-1}(\vec{p},c))\times (a,b) $ is a homeomorphism
for every connected component  $Z$ of a stratum of $\mathcal{Z}$.
\end{enumerate}
This statement of Thom's First Isotopy Lemma is a specialized form
of Theorem {II}.6.2 in Shiota~\cite{Shiota}.

Remark that condition (a) is automatically satisfied. Indeed, the
inverse image by $f_{\vec{p}}$ of any interval $[a,b]\subset\Rset$
is equal to $\clo{A}\cap (B^n(\vec{p},b)-\inte{B^n(\vec{p},a)})$
which is closed and bounded in $\Rset^n$.

\begin{claim}\label{claim:int}
If condition (b) is satisfied for $0<b$ (so $a=0$), then
every $c\in(0,b)$ is a cone radius of $A$ in $\vec{p}$.
\end{claim}
\begin{proof}[Proof of Claim] Take an arbitrary real
number $c\in(0,b)$. The lemma gives a homeomorphism
$$
h_0:\clo{A}\cap\inte{B^n(\vec{p},b)-\{\vec{p}\}}\rightarrow(\clo{A}\cap S^{n-1}(\vec{p},c))\times
(0,b).
$$
By  property (i), we obtain a
homeomorphism
$$
h_1:\clo{A}\cap (B^n(\vec{p},c)-\{\vec{p}\})\rightarrow(\clo{A}\cap S^{n-1}(\vec{p},c))\times
(0,c],
$$
which equals the restriction $h_0|\clo{A}\cap
(B^n(\vec{p},c)-\{\vec{p}\})$.
Since the cylinder
$(\clo{A}\cap S^{n-1}(\vec{p},c))\times(0,c]$ is clearly homeomorphic to
the cone
$\mathrm{Cone}(\clo{A}\cap S^{n-1}(\vec{p},c),\vec{p})-\{\vec{p}\}$, e.g., by the
homeomorphism
$$
h_2(\vec{x},t):=(1-\frac{t}{c})\vec{p}+(\frac{t}{c})\vec{x}\quad\text{for}\,
t\in(0,c],
$$
we obtain a homeomorphism
$$
h_3:=h_2\circ h_1:\clo{A}\cap (B^n(\vec{p},c)-\{\vec{p}\})\rightarrow
\mathrm{Cone}(\clo{A}\cap S^{n-1}(\vec{p},c),\vec{p})-\{\vec{p}\}.
$$

It is easily verified that $h_2:(Z\cap
S^{n-1}(\vec{p},c))\times(0,c]\rightarrow \mathrm{Cone}(Z\cap
S^{n-1}(\vec{p},c),\vec{p})-\{\vec{p}\}$ is a homeomorphism for each connected
component $Z$ of a stratum of $\mathcal{Z}$. Since $h_1$ also satisfies property
(ii), we have that
$h_3:(Z\cap (B^n(\vec{p},c)-\{\vec{p}\})\rightarrow \mathrm{Cone}(Z\cap
S^{n-1}(\vec{p},c),\vec{p})-\{\vec{p}\}$ is  a homeomorphism for each connected component
$Z$ of a stratum of $\mathcal{Z}$.

This implies that the restriction $h=h_3| A\cap
 (B^n(\vec{p},c)-\{\vec{p}\})$ is a homeomorphism
from  $h(A\cap (B^n(\vec{p},c)-\{\vec{p}\})$ to $\mathrm{Cone}(A\cap
S^{n-1}(\vec{p},c),\vec{p})-\{\vec{p}\}$, because $A$ is the union
of connected components of strata of $\mathcal{Z}$.

The homeomorphism $h$ can easily be
extended to the point $\vec{p}$, hence $c$ is indeed a cone radius
as desired.
\end{proof}

Let $\mathcal{S}$ be a schema containing a relation name $S$ of arity $n$, and let $D$ be a database
over $\mathcal{S}$.
By Claim~\ref{claim:int}, we can define the following cone radius query
$$
Q_{\text{radius}}(D):=\{(\vec{p},r)\in\Rset^{n}\times\Rset\mid
\vec{p}\in S^D  \text{ and } r\in (0,b)\},
$$
where the interval $(0,b)$ satisfies condition (b) for the map $f_{\vec{p}}$ 
 and semi-algebraic set
$A=S^D$. Let us express this query in FO\@.

We  define the critical point query as
\begin{multline*}
Q_{\text{crit}}(D):=\{
(\vec{p},\vec{x})\in\Rset^n\times\Rset^n\mid \vec{p}\in S^D \text{ and }
\vec{x}\in Q_{\text{$k$-stratum}}(D)
\text{ for a certain } k \\
\text{ and } \vec{x} \text{ is a critical point of }
f_{\vec{p}} \text{ restricted to } Q_{\text{$k$-stratum}}(D)\}.
\end{multline*}

The \textit{critical points} of $f_{\vec{p}}$ restricted to a stratum $Z$,
are the points $\vec{x}\in Z$ where
 the differential map $d_{\vec{x}}(f_{\vec{p}}|Z)$
is not surjective.
\begin{claim}
A point $\vec{x}\in\Rset^n$ is a critical point
of $f_{\vec{p}}|Z$ if and only if the tangent space
of $Z$ in $\vec{x}$ is orthogonal to the vector $\vec{x}-\vec{p}$.
\end{claim}
\begin{proof}[Proof of Claim]
 We compute the differential $d_{\vec{x}}(f_{\vec{p}}|Z)$ as follows:
Locally around $\vec{x}$, we may assume that the
projection on the first $k$ coordinates
$\Pi:Z\rightarrow U\subset\Rset^k$, is a homeomorphism,
 where $k$ is the dimension of $Z$ in
$\vec{x}$. By definition of the differential,
$
d_{\vec{x}}(f_{\vec{p}}|Z)=(d_{\tup{x}{k}}g) (d_{\tup{x}{k}}\Pi^{-1})^{-1},
$
where $g=(f|Z)\circ\Pi^{-1}$.
By the $C^1$ Inverse Function Theorem, we may assume that
 $\Pi^{-1}:U\rightarrow
 Z:\tup{x}{k}\mapsto(\tupp{x}{k},\tupp[k+1]{\varphi}{n})$,
 where $\varphi_{i}\tup{x}{k}$ are $C^1$-maps, and hence
 $
 g:U\mapsto\Rset:\tup{x}{k}=\sum_{i=1}^k (x_i-p_i)^2 +
 \sum_{j=k+1}^n (\varphi_j\tup{x}{k}-p_j)^2.
 $
An elementary calculation shows that the differential of
$f_{\vec{p}}|Z$ in $\vec{x}$ is the vector $$
d_{\vec{x}}(f_{\vec{p}}|Z)=2\left(\left(
(x_i-p_i)+\sum_{j=k+1}^n(x_j-p_j)\frac{\partial\varphi_j}{\partial
x_i}\tup{x}{k}\right)_{i=1,\ldots,k},\underbrace{0,\ldots,0}_{\text{$n-k$
times}}\right). $$ Since $d_{\tup{x}{k}}\Pi^{-1}$ is an
isomorphism between the tangent space $T_{\tup{x}{k}}U$ of $U$ in
the projection $\Pi(\vec{x})$, and the tangent space
$T_{\vec{x}}Z$ of $Z$ in $\vec{x}$, any tangent vector
$\tup{v}{n}\in T_{\vec{x}}Z$ is of the form
$(d_{\tup{x}{k}}\Pi^{-1})\tup{v}{k}$.
More specifically, any tangent vector $\vec{v}\in T_{\vec{x}}Z$
can be written as $$
\tup{v}{n}=(\tupp{v}{k},\sum_{i=1}^k\frac{\partial\varphi_{k+1}}{\partial
x_i}\tup{x}{k}v_i,\ldots,\sum_{i=1}^k\frac{\partial\varphi_n}{\partial
x_i}\tup{x}{k}v_i). $$ Hence, the product
$$
d_{\vec{x}}(f_{\vec{p}}|Z)\cdot\vec{v}=
2\sum_{i=1}^{k}(x_i-p_i)v_i +2
\sum_{j=k+1}^n (x_j-p_j)
\left(\sum_{i=1}^k\frac{\partial\varphi_{j}}{\partial
x_i}\tup{x}{k}v_i \right),
$$
 is equal to
$2\sum_{i=1}^n (x_i-p_i)v_i$.
This implies that the differential map $d_{\vec{x}}(f_{\vec{p}}|Z)$ is not
surjective if and only if $2\sum_{i=1}^n (x_i-p_i)v_i=0$ for all
tangent vectors $\vec{v}\in T_{\vec{x}}Z$. This proves the Claim.
\end{proof}

The proof of the theorem now continues as follows.
The tangent space query
$$
Q_{\text{tangent}}(D):=\{(\vec{x},\vec{v})
\in\Rset^n\times\Rset^n\mid \text{$S^D$ is $C^1$,\, $\vec{x}\in S^D$ 
 and } \vec{v}\in T_{\vec{x}}S^D\},
$$
is expressible in FO~\cite[Lemma 2]{Rannou}.
Because the orthogonality of two vectors can be easily expressed in
FO, the formula
$$
\varphi_{\text{crit}}(\vec{p},\vec{x}):=S(\vec{p})\land
\bigvee_{j=0}^n \forall
\vec{v}\left(
\varphi_{\text{tangent}}(\varphi_{\text{$j$-stratum}}(S))(\vec{x},\vec{v}) \rightarrow
(\vec{x}-\vec{p})\cdot\vec{v}=0
\right)
$$
expresses $Q_{\text{crit}}$ correctly by Claim 2. Here,
 $\varphi_{\text{$j$-stratum}}$ denotes an FO-formula expressing
  $Q_{\text{$j$-stratum}}$ for $j=0,\ldots,n$, and
 $\varphi_{\text{tangent}}$ is an  FO formula expressing $Q_{\text{tangent}}$.

A \textit{critical value} of $f_{\vec{p}}$ is the image by
$f_{\vec{p}}$ of a critical point. The query which returns
 the set of critical values is expressible in FO by the formula
$$
\varphi_{\text{val}}(\vec{p},r):=\exists \vec{x}\left(
\varphi_{\text{crit}}(\vec{p},\vec{x}) \land  r=f_{\vec{p}}(\vec{x})\right).
$$

We now observe that  $\{r\in\Rset\mid
\varphi_{\text{val}}(\vec{p},r)\}$ is finite  for each $\vec{p}$.
Indeed, the set of critical points
$\{\vec{x}\in\Rset^n\mid \varphi_{\text{crit}}(\vec{p},\vec{x})\}$
 is semi-algebraic and hence admits a $C^1$-cell decomposition
  $\mathcal{C}=\{C_1,\ldots,C_m\}$ such
that $f|C_i$ is $C^1$~\cite{Dries}. Sard's Theorem for
$C^1$-maps~\cite{wilkie} implies that each $f_{\vec{p}}|C_i$ attains only a
finite number of values. Hence the image by $f_{\vec{p}}$ of
the set of critical points is finite.

This implies that either there are no critical values, or there
exists a minimal critical value. By Claim 1, any value smaller
than this minimal value is a cone radius. We therefore conclude
that the query expressed in FO\ as
$$
\varphi_{\text{radius}}(\vec{p},r):=(\forall r'
)(\varphi_{\text{val}}(\vec{p},r')\rightarrow r<r'),
$$
is a cone radius query, as desired.
\end{proof}
\subsection*{Acknowledgement} The author wishes to thank Jan Van den Bussche
 for many interesting discussions and critical remarks, which  led
 to improvements in the presentation of the paper.

\end{document}